\documentclass{article}%
\usepackage{amsfonts}
\usepackage{aip}
\usepackage{amsmath}
\usepackage{amssymb}
\usepackage{graphicx}%
\providecommand{\U}[1]{\protect\rule{.1in}{.1in}}

\begin{document}

\title{Eigenfunctions for Liouville Operators, Classical Collision Operators, and
Collision Bracket Integrals in Kinetic Theory}
\author{Byung Chan Eu\\Department of Chemistry, McGill University\\801 Sherbrooke St. West, Montreal\\Quebec H3A 2K6, Canada}
\maketitle

\begin{abstract}
In the kinetic theory of dense fluids the many-particle collision bracket
integral is given in terms of a classical collision operator defined in the
phase space. To find an algorithm to compute the collision bracket integrals,
we revisit the eigenvalue problem of the Liouville operator and re-examine the
method previously reported[Chem. Phys. \textbf{20}, 93(1977)]. Then we apply
the notion and concept of the eigenfunctions of the Liouville operator and
knowledge acquired in the study of the eigenfunctions to obtain alternative
forms for collision integrals. One of the alternative forms is given in the
form of time correlation function. This form, on an additional approximation,
assumes a form reminiscent of the Chapman-Enskog collision bracket integral
for dilute gases. It indeed gives rise to the latter in the case of two
particles. The alternative forms obtained are more readily amenable to
numerical simulation methods than the collision bracket integras expressed in
terms of a classical collision operator, which requires solution of classical
Lippmann-Schwinger integral equations. This way, the aforementioned kinetic
theory of dense fluids is made more accessible by numerical
computation/simulation methods than before.

\end{abstract}

\section{Introduction}%

\setlength{\baselineskip}{20pt}%
Classical collision operators\cite{prigogine,zwanzig,eubook} defined in the
phase space are classical mechanical analogs of quantum mechanical collision
operators defined in terms of Liouville--von Neumann operators and the
classical limits of the latter. They appear in the statistical mechanical
formulas of transport coefficients in kinetic theory of matter and present an
important problem to resolve in the final stage of implementing the dense
fluid kinetic theory to make it connect with experiments. Although practical
methods of computing the classical collision operators are essential to making
kinetic theory a useful molecular theory of matter, they have not been given
much attention beyond the formal theory level in contrast to their quantum
mechanical analogs in the literature\cite{newton,faddeev,prugovecki}, and
computation of them and related expressions appearing in kinetic theory still
poses a theoretical challenge in nonequilibrium statistical mechanics. In this
paper, we would like to take up the subject for study and make an attempt to
mitigate the situation.

Classical collision operators are generally defined in formal
analogy\cite{miles,eu71} to quantum mechanical collision operators in Hilbert
space, but their definitions have been made without much attention paid to the
space of functions in which the operators live. Clearly, study of classical
collision operators would require a space of eigenfunctions for the underlying
Liouville operator in the phase space. However, the eigenfunctions of the
Liouville operator are little known except for free particles. Here we first
consider the classical eigenvalue problems for Liouville operators, especially
when the spectrum is continuous, and then apply the eigenfunctions to recast
many-particle collision bracket integrals, which are given in terms of
classical collision operators and appear in the kinetic theory of dense
fluids. In Ref. 9, the present author made a study of eigenvalue problems for
Liouville operators from the prespective of scattering theory of few particle
systems. In the present paper, we revisit the problem and improve upon the
method of calculating eigenfunctions before applying the notion and existence
of eigenfunctions to examine a way to calculate the collision bracket
integrals in the context of the kinetic theory of transport processes,
particularly, in liquids. We will examine the eigenfunctions for classical
scattering of particles in phase space and make use of the acquired results to
cast the collision bracket integral given in terms of a classical collision
operator into a form more readily amenable to numerical computation/simulation
methods. Since the many-body problems involved in such a theory do not allow
simple analytic forms for the collision cross sections and transition
probabilities for physically realistic dynamical events and hence pose a
barrier to overcome and thereby make it practicable in kinetic theory in
general, it is imperative to devise alternative forms of transport
coefficients expressed in terms of classical collision operators, so that
efficient numerical algorithms can be found for them. Therefore, this line of
study is quite relevant to implementing the classical kinetic theory of matter
to understand transport properties of dense gases and liquids studied in the laboratory.

In Section \ref{Sec2}, we re-examine the eigenvalue problem for classical
Liouville operator and obtain the eigenfunctions in terms of a generating
function for canonical transformation in a more concise manner than in the
previous work\cite{eu77}. The associated scattering theory in phase space is
considered in Section \ref{Sec3}. The results, at least the basic concept,
obtained in Section \ref{Sec2} will be used to cast the classical collision
operator in a numerically convenient form in Section \ref{Sec4}. The new form
of the collision bracket integral will be shown to yield its low density
limiting form in the same form as the Chapman-Enskog results for the transport
coefficients for dilute gases. A form of collision bracket integrals involving
a three-body collision operator is also presented to indicate the general idea
of the method in the case of many-particle situations. Concluding remarks are
given in Section \ref{Sec5}.

\section{The Eigenvalue Problem for the Classical Liouville
Operator\label{Sec2}}

In the kinetic theory of matter\cite{zwanzig,eubook,snider,cohen,kawasaki},
the formalism is usually formulated in terms of the Liouville operator, and
associated classical collision operators appear in close analogy to quantum
scattering theory. Formal theory\cite{newton} of quantum scattering has been
studied in depth from the mathematical standpoint and the mathematical
properties\cite{faddeev,prugovecki} of quantum scattering operators, such as
the Hilbert space for the operators, are well understood at present. On the
other hand, it cannot be said the same of the classical scattering theory
based on the classical Liouville operator, and even the mathematical nature of
the function space for the classical Liouville operator is not well clarified
at present, almost all of mathematical operations involving the classical
Liouville operators being performed by simple analogy to the quantum
counterparts; see, for example, Refs. 2, 3, 7--11. Here in this section we
will consider an aspect of the problem, limiting our study to construction of
eigenfunctions, which may be given in terms of the Hamilton--Jacobi
characteristic function\cite{goldstein}. The eigenfunctions are $L^{2}$
functions normalizable to unity and have a closure in phase space. If the
Hamilton--Jacobi equation is separable, then the eigenfunctions can be given
in terms of quadratures. Otherwise, a numerical solution method is the only
way left to treat the problem adequately. Nevertheless, the eigenfunctions can
be used to formulate formal classical scattering theory in analogy to quantum
scattering theory, and the theory thus formulated can provide a mathematically
transparent and practicable computational algorithm for the classical
collision problems associated with transport coefficients of gases and liquids.

The particles are assumed to interact by the pair potentials $V_{jk}%
(\mathbf{r}_{jk})$, where $\mathbf{r}_{jk}$ is the relative coordinate vector
between the two particles $j$ and $k$. Let $\mathbf{p\equiv}\left(
p_{1},\cdots,p_{f}\right)  $\textbf{ }and\textbf{ }$\mathbf{r\equiv}\left(
r_{1},\cdots,r_{f}\right)  $\ denote momenta and positions of particles with
$f$ standing for the number of degrees of freedom. The Hamiltonian in the
relative coordinate system will be denoted by $H\left(  \mathbf{p,r}\right)
$:%
\begin{equation}
H\left(  \mathbf{p,r}\right)  =\sum_{j}\frac{p_{j}^{2}}{2m_{j}}+\sum
_{j<k}V_{jk}(\mathbf{r}_{jk}) \label{hamilton}%
\end{equation}
under the assumption that the interaction potential energies are pairwise
additive. The Liouville operator is then defined by the Poisson brackets times
$-i$ where $i=\sqrt{-1}$:
\begin{equation}
\mathbf{L}=-i\left[  H,\;\right]  _{\mathbf{pr}}=-i\sum_{k}\left(
\frac{\partial H}{\partial p_{k}}\frac{\partial}{\partial r_{k}}%
-\frac{\partial H}{\partial r_{k}}\frac{\partial}{\partial p_{k}}\right)  .
\label{1}%
\end{equation}
With $-i$ multiplied, the Liouville operator becomes self-adjoint. The
subscript $k$ denotes Cartesian components of vectors $\mathbf{p}$ and
$\mathbf{r}$. With this definition of Liouville operator the Liouville
equation can be written as
\begin{equation}
i\frac{\partial\rho}{\partial t}=\mathbf{L}\rho(x,t), \label{2}%
\end{equation}
where $x$ is the abbreviation for the phase $x=\left(  \mathbf{p,r}\right)  $
and $\rho$ is the probability distribution function. Since it is possible to
expand $\rho$ into Fourier components as in the integral
\begin{equation}
\rho(x,t)=\int_{-\infty}^{\infty}d\lambda\exp\left(  -i\lambda t\right)
\psi_{\lambda}(x), \label{3}%
\end{equation}
we arrive at an eigenvalue problem\cite{prigogine} of $\mathbf{L}:$%
\begin{equation}
\mathbf{L}\left(  x\right)  \psi_{\lambda}(x)=\lambda\psi_{\lambda}(x),
\label{4}%
\end{equation}
where $\lambda$ is an eigenvalue and $\psi_{\lambda}$ the eigenfunction
belonging to $\lambda$. This eigenvalue problem is subject to appropriate
boundary conditions. It is intimately related to the Hamilton--Jacobi theory
in classical mechanics as will be shown below.

Let us consider the eigenvalue problem, Eq. (\ref{4}), in another form, which
appears to be more insightful. The Liouville operator $\mathbf{L}$ is
Hermitian in the space of square-integrable functions $\psi_{\lambda}$, which
are generally complex. Therefore, if $\psi_{\lambda}(x)$ is written in the
form
\begin{equation}
\psi_{\lambda}(x)=A(x)\exp\left[  i\Gamma\left(  x\right)  \right]  ,
\label{8}%
\end{equation}
then the eigenvalue problem can be cast in the pair of equations
\begin{equation}
\left[  H,\Gamma\right]  _{\mathbf{pr}}=\lambda, \label{9}%
\end{equation}%
\begin{equation}
\left[  H,A\right]  _{\mathbf{pr}}=0. \label{10}%
\end{equation}
Eq. (\ref{10}) indicates that the amplitude $A(x)$ is a function of canonical
invariants. Therefore the amplitude $A(x)$ is a function of the basic
invariants of $H$; for exapmple, a function of the Hamiltonian and the total
momentum. However, the precise form for the function is not obvious at this
point of development.

Since the Liouville operator is invariant to canonical transformation $\left(
\mathbf{p,r}\right)  \rightarrow\left(  \mathbf{P,Q}\right)  $ where
$\mathbf{P}$ and $\mathbf{Q}$ are the new canonical momentum and coordinate
preserving the forms of Hamilton's canonical equations of motion, in the new
canonical variable system the eigenvalue problem takes the form
\begin{equation}
\mathbf{L}\left(  X\right)  \psi_{\lambda}(X)=\lambda\psi_{\lambda}(X),
\label{5}%
\end{equation}
where $X=\left(  \mathbf{P,Q}\right)  $ and
\begin{equation}
\mathbf{L}\left(  X\right)  =-i\left[  H,\;\right]  _{\mathbf{PQ}}=-i\sum
_{k}\left(  \frac{\partial H}{\partial P_{k}}\frac{\partial}{\partial Q_{k}%
}-\frac{\partial H}{\partial Q_{k}}\frac{\partial}{\partial P_{k}}\right)  .
\label{6}%
\end{equation}
Here $H$ is the new Hamiltonian $H=H\left(  \mathbf{P,Q}\right)  $. If the new
Hamiltonian is independent of $\mathbf{Q}$, then we have
\begin{equation}
\mathbf{L}\left(  X\right)  =-i\left[  H,\;\right]  _{\mathbf{PQ}}=-i\sum
_{k}\frac{\partial H}{\partial P_{k}}\frac{\partial}{\partial Q_{k}}.
\label{7}%
\end{equation}
We will return to this form for $\mathbf{L}$ presently. Other mathematical
properties of $\mathbf{L}$ and aspects of the eigenvalue problem in hand are
discussed in Appendix A. They are collected in Appendix $A$ for completeness
and also to make this article self-contained.

Let us assume $S(\mathbf{P,r},t)$ to be the generating
function\cite{goldstein} of canonical transformation such that
\begin{equation}
p_{k}=\frac{\partial S}{\partial r_{k}},\quad Q_{k}=\frac{\partial S}{\partial
P_{k}}. \label{11}%
\end{equation}
The generating function $S$ obeys the Hamilton--Jacobi equation
\begin{equation}
\frac{\partial S}{\partial t}+H\left(  \frac{\partial S}{\partial\mathbf{r}%
},\mathbf{r}\right)  =0. \label{12}%
\end{equation}
Under this canonical transformation the amplitude function is given by
\begin{equation}
A(x)=\delta\left(
\mbox{\boldmath$\alpha$}%
-\frac{\partial S}{\partial\mathbf{r}}\right)  , \label{13}%
\end{equation}
where $\alpha_{1},\cdots,\alpha_{f}$ are the values of $P_{1},\cdots,P_{f}$,
respectively, with $f$ denoting the degrees of freedom. It is sufficient to
take this form because the Hamilton--Jacobi equation is subject to a constant
energy for a conservative system. It is easily shown that the expression for
$A$ given above satisfies Eq. (\ref{10}).

To find the phase function $\Gamma$, we observe that Eq. (\ref{9}), on
canonical transformation, can be written as
\begin{equation}
\left[  H,\Gamma\right]  _{\mathbf{PQ}}=\lambda. \label{14}%
\end{equation}
Explicitly written out, Eq. (\ref{14}) has the form
\begin{equation}
\lambda=\sum_{k}\frac{\partial H}{\partial P_{k}}\frac{\partial\Gamma
}{\partial Q_{k}}. \label{14'}%
\end{equation}
It is useful to define
\begin{equation}
\omega_{k}=\frac{\partial H}{\partial P_{k}}, \label{15}%
\end{equation}
which is independent of $\mathbf{Q}$ since $H=H\left(  \mathbf{P}\right)  $ is
independent of $\mathbf{Q}$. Then, Eq. (\ref{14'}) is readily solved, and we
find
\begin{equation}
\Gamma=\sum_{k}\frac{\lambda_{k}}{\omega_{k}}Q_{k}+\Gamma_{0}, \label{16}%
\end{equation}
where $\Gamma_{0}$ is a constant, which can be absorbed into the normalization
factor, and constants (eigenvalues) $\lambda_{k}$ are such that
\[
\lambda=\sum_{k}\lambda_{k}.
\]
They will be more precisely determined on imposing suitable boundary
conditions on the eigenfunctions $\psi_{\lambda}$. It should be noted that
$Q_{k}=Q_{k}\left(  \mathbf{r},%
\mbox{\boldmath$\alpha$}%
\right)  $.

In summary of the results up to this point, under the canonical transformation
(\ref{11}) the eigenfunctions are in the form
\begin{align}
\psi_{\lambda}  &  =A_{0}\delta\left(
\mbox{\boldmath$\alpha$}%
-\frac{\partial S}{\partial\mathbf{r}}\right)  \exp\left(  i\sum_{k}%
\frac{\lambda_{k}}{\omega_{k}}\frac{\partial S}{\partial\alpha_{k}}\right)
\nonumber\\
&  =A_{0}\delta\left(
\mbox{\boldmath$\alpha$}%
-\mathbf{p}\right)  \exp\left(  i\sum_{k}\frac{\lambda_{k}}{\omega_{k}}%
Q_{k}\right)  , \label{17}%
\end{align}
where $A_{0}$ is the normalization factor, and $p_{k}$ and $Q_{k}$ are used
for the second equality of Eq. (\ref{17}). Since the generating function $S$
can be written for a conservative system in terms of the characteristic
function $W$ defined by
\begin{equation}
S=W-Et, \label{18}%
\end{equation}
where $E$ is a constant (energy), the eigenfunction can be also written as
\begin{equation}
\psi_{\lambda}=A_{0}\delta\left(
\mbox{\boldmath$\alpha$}%
-\mathbf{p}\right)  \exp\left(  i\sum_{k}\frac{\lambda_{k}}{\omega_{k}}%
\frac{\partial W}{\partial\alpha_{k}}\right)  . \label{17'}%
\end{equation}
This result shows that the eigenfunction will be explicitly found if the
Hamilton--Jacobi equation is solved, given the initial conditions for $\left(
\mathbf{p,r}\right)  $.

This function $\psi_{\lambda}=\psi\left(  \mathbf{r,p;}%
\mbox{\boldmath$\alpha$}%
,%
\mbox{\boldmath$\lambda$}%
\right)  $ may be regarded as an eigenfunction with two sets of eigenvalues
$\left(
\mbox{\boldmath$\alpha$}%
,%
\mbox{\boldmath$\lambda$}%
\right)  $. This set of eigenfunctions can be easily shown orthogonal and
normalizable. For this proof we simply note that under the canonical
trasformation and subject to $\delta\left(
\mbox{\boldmath$\alpha$}%
-\mathbf{p}\right)  $
\[
\frac{\partial\left(  Q_{1},\cdots,Q_{f}\right)  }{\partial\left(
r_{1},\cdots,r_{f}\right)  }=\frac{\partial\left(  p_{1},\cdots p_{f}\right)
}{\partial\left(  \alpha_{1},\cdots,\alpha_{f}\right)  }=1.
\]

Before proceeding further, it is interesting to note the following. Recalling
that the generating function $S$ can be expressed in terms of the action
integral%
\begin{equation}
S(t)=-\int_{t_{0}}^{t}ds\mathcal{L}\left(  \mathbf{p},\mathbf{r},s\right)  ,
\label{17d}%
\end{equation}
where $\mathcal{L}\left(  \mathbf{p},\mathbf{r},s\right)  $ is the Lagrangian,
the eigenfunctions, e.g., Eq. (\ref{17c}), may be expressed in the form%
\begin{align}
\psi\left(  \mathbf{r,p;}%
\mbox{\boldmath$\alpha$}%
,%
\mbox{\boldmath$\lambda$}%
\right)   &  =\frac{A_{0}}{\left(  2\pi\right)  ^{3f}}\delta\left(
\mbox{\boldmath$\alpha$}%
-\frac{\partial}{\partial\mathbf{r}}\int_{t_{0}}^{t}ds\mathcal{L}\left(
\mathbf{p},\mathbf{r},s\right)  \right)  \times\nonumber\\
&  \exp\left[  -i\sum_{j}\frac{\lambda_{j}}{\omega_{j}}\frac{\partial
}{\partial\alpha_{j}}\int_{t_{0}}^{t}ds\mathcal{L}\left(  \mathbf{p}%
,\mathbf{r},s\right)  \right]  , \label{17e}%
\end{align}
which reminds us of the path integral\cite{feynman} in quantum mechanics in
that it is given in terms of the action integral. It is useful to note%
\begin{equation}
\int_{t_{0}}^{t}ds\mathcal{L}\left(  \mathbf{p},\mathbf{r},s\right)  =\sum
_{j}\int_{t_{0}}^{t}dsp_{j}\frac{dr_{j}}{ds}-E\left(  t-t_{0}\right)
=\sum_{j}\int_{r_{j0}}^{r_{j}}dr_{j}p_{j}-E\left(  t-t_{0}\right)  .
\label{17f}%
\end{equation}

We now consider periodic boundary conditions on the eigenfunctions. Let
$\mathbf{\Omega}$ be the dimension of the cubic box. Then the eigenfunctions
must obey the condition
\begin{equation}
\psi_{\lambda}(\mathbf{r}+\mathbf{\Omega)}=\psi_{\lambda}(\mathbf{r).}
\label{19}%
\end{equation}
It follows
\begin{equation}
\frac{\lambda_{j}}{\omega_{j}}\left[  Q_{j}\left(  \mathbf{r}+\mathbf{\Omega
}\right)  -Q_{j}\left(  \mathbf{r}\right)  \right]  =2\pi l_{j}. \label{20}%
\end{equation}
Here $l_{j}$ is an integer. Since $Q_{j}\left(  \mathbf{r}+\mathbf{\Omega
}\right)  -Q_{j}\left(  \mathbf{r}\right)  =\Omega$ it follows that%
\begin{equation}
\lambda_{j}=\frac{2\pi l_{j}\omega_{j}}{\Omega}. \label{20a}%
\end{equation}
Define the wave number $k_{j}$
\begin{equation}
k_{j}=\frac{2\pi l_{j}}{\Omega}\quad\left(  l_{j}=0,\pm1,\pm2,\cdots\right)  .
\label{23}%
\end{equation}
Hence the eigenfunctions are:
\begin{equation}
\psi\left(  \mathbf{r,p;}%
\mbox{\boldmath$\alpha$}%
,%
\mbox{\boldmath$\lambda$}%
\right)  =\frac{A_{0}}{\left(  2\pi\right)  ^{3f}}\delta\left(
\mbox{\boldmath$\alpha$}%
-\mathbf{p}\right)  \exp\left(  i\sum_{j}k_{j}Q_{j}\right)  . \label{23a}%
\end{equation}

Since the characteristic function $W$ is a surface in the phase space and the
trajectories of $l_{j}\neq0$ are on the family of surfaces, the eigenfunctions
represent waves propagating perpendicularly to the surface of the
characteristic function $W$ with the phase
\begin{equation}
\Gamma=\sum_{j}\frac{\lambda_{j}}{\omega_{j}}\frac{\partial W}{\partial
\alpha_{j}}=\sum_{j}k_{j}\frac{\partial W}{\partial\alpha_{j}}=\sum_{j}%
k_{j}Q_{j}\left(  \mathbf{r},%
\mbox{\boldmath$\alpha$}%
\right)  . \label{24}%
\end{equation}

Eqs. (\ref{17'}) and (\ref{24}) represent the formal solution of the
eigenvalue problem in terms of the Hamilton--Jacobi characteristic function
$W$. This method of constructing the eigenfunctions is not only slightly
different from, but also more insightful than that obtained in the previous
work\cite{eu77} by the present author.

For noninteracting particles we easily obtain
\begin{equation}
\psi_{\lambda}=A_{0}\delta\left(
\mbox{\boldmath$\alpha$}%
-\mathbf{p}\right)  \exp\left(  i\sum_{j}k_{j}r_{j}\right)  , \label{25}%
\end{equation}
in agreement with the known result in the literature\cite{prigogine}.

It is an interesting exercise to construct the eigenfunctions for separable
bound-state problems in the case of two interacting particles. Such examples
are given for a few cases in Ref. 9. Such eigenfunctions can be used for
calculations of few-body dynamics problems related to bound states or
scattering of isolated two-particle systems in a classical approximation to
quantum dynamics.

\section{Classical Scattering Theory in Phase Space\label{Sec3}}

Collision bracket integrals appear in the formulas\cite{eubook,eubk2,eubk4} of
transport processes when dynamic processes are treated for gases and liquids
in kinetic theory. If we wish to achieve a molecular theory of such processes
it is essential to calculate them in terms of molecular information on the
basis of mechanical laws. Since it wouldn't be possible to expect to evaluate
them in analytic form for a realistic potential model, it would be essential
to develop numerical computational algorithms for them.

As a preparation for expressing collision bracket integrals involving
classical collision operators in an equivalent but computationally more
practical form and also to make this work self-contained, we briefly review
classical scattering theory in the phase space. Formal classical theory of
scattering\cite{prigogine,eubook,miles,eu77} can be formulated in the phase
space in a manner parallel to the quatum mechanical scattering theory, and it
holds some advantages for statistical mechanics and, especially, for kinetic
theory as has been frequently demonstrated in kinetic theory
investigations\cite{zwanzig,eubook,cohen,kawasaki} in which classical
collision operators are used in a formalism analogous to quantum scattering
theory. Nevertheless, the meanings of such collision operators have not been
studied beyond the formal definition level. Since they are often used in such
investigations but wihtout their computational methods sufficiently well
discussed, it is all the more important to try to comprehend their
mathematical basis, so that one can perform computations, for example, of the
collision integrals appearing in the theory of transport processes. Our aim
here, however, is not in developing classical scattering theory in depth, but
rather in exposing the essential features that may be relevant to mathematical
treatments of and developing computational algorithms and suitable
approximations for them which one might use in future calculations of
transport properties. For this purpose the eigenfunctions presented in the
previous section provide valuable mathematical tools and a conceptual
framework as well as a methodology for such efforts.

Consider a scattering situation where particles (beams) at infinite separation
converge toward each other and interact and then separate to infinite relative
distance from each other. In classical mechanics, there is no concept of waves
for particle motion. However, when the collision problem is formulated in the
phase space, there appears a notion of waves as we have seen in the previous
section where the eigenfunctions are calculated for continuous spectra of the
Liouville operator. The Liouville equation (\ref{2}) governs this scattering
process of waves in the phase space. We describe the basic aspects of the
theory here.

The collision process of the particles is assumed governed by the Liouville
equation (\ref{2}) over the course of collision. Since at remote past and
distant future where particles do not interact with each other, the
probability distribution function $\rho$ also obeys the free particle
Liouville equation
\begin{equation}
i\frac{\partial\rho}{\partial t}=\mathbf{L}_{0}\rho, \label{26}%
\end{equation}
where $L_{0}$ is the free Liouville operator defined by
\begin{equation}
\mathbf{L}_{0}=-i\sum_{k}\frac{p_{k}}{m}\frac{\partial}{\partial r_{k}}.
\label{26a}%
\end{equation}
If the system is prepared at an eigenstate $\lambda$ at remote past and thus
we set
\begin{equation}
\rho\left(  x,t\right)  =\exp\left(  -i\lambda t\right)  \Phi_{\lambda}\left(
x\right)  , \label{27}%
\end{equation}
the corresponding eigenvalue problem is
\begin{equation}
\mathbf{L}_{0}\Phi_{\lambda}\left(  x\right)  =\lambda\Phi_{\lambda}\left(
x\right)  . \label{28}%
\end{equation}
This eigenfunction is of the form as in Eq. (\ref{25}). To describe the
scattering process it is useful to introduce the interaction representation.
Therefore, we define
\begin{equation}
\tilde{\rho}(t)=\exp\left(  i\mathbf{L}_{0}t\right)  \rho\left(  t\right)  .
\label{29}%
\end{equation}
This puts the Liouville equation (\ref{2}) in the form
\begin{equation}
i\frac{\partial\tilde{\rho}}{\partial t}=\mathbf{\tilde{L}}_{1}(t)\tilde{\rho
}(t), \label{30}%
\end{equation}
where
\begin{equation}
\mathbf{\tilde{L}}_{1}(t)=\exp(i\mathbf{L}_{0}t)\mathbf{L}_{1}\exp
(-i\mathbf{L}_{0}t), \label{31}%
\end{equation}
with $\mathbf{L}_{1}$ denoting the interaction Liouville operator
\begin{equation}
\mathbf{L}_{1}=i\sum_{k}\frac{\partial H}{\partial r_{k}}\frac{\partial
}{\partial p_{k}}. \label{31a}%
\end{equation}
The formal solution for Eq. (\ref{30}) can be written in the form
\begin{equation}
\tilde{\rho}(t)=\exp(i\mathbf{L}_{0}t)\exp\left[  -i\mathbf{L}\left(
t-t_{0}\right)  \right]  \exp(-i\mathbf{L}_{0}t)\tilde{\rho}(t_{0}).
\label{32}%
\end{equation}
As $t_{0}\rightarrow-\infty$, where the interaction between the particles
vanishes, there holds the limit $\mathbf{L}\rightarrow\mathbf{L}_{0}$ and
therefore
\begin{equation}
\tilde{\rho}(t)\rightarrow\exp(i\mathbf{L}_{0}t)\exp(-i\mathbf{L}_{0}%
t)\Phi_{\lambda}=\Phi_{\lambda}, \label{33}%
\end{equation}
and similarly as $t\rightarrow\infty$. Consequently, in the limit of
$t_{0}\rightarrow-\infty$ the distribution function is given by
\begin{align}
\rho(t)  &  =\lim_{t_{0}\rightarrow-\infty}\exp\left[  -i\mathbf{L}\left(
t-t_{0}\right)  \right]  \exp(-i\mathbf{L}_{0}t_{0})\tilde{\rho}%
(t_{0})\nonumber\\
&  =\lim_{t_{0}\rightarrow-\infty}\exp\left[  -i\mathbf{L}\left(
t-t_{0}\right)  \right]  \exp(-i\mathbf{L}_{0}t_{0})\Phi_{\lambda}. \label{34}%
\end{align}
This is a strong limit which may be written in an equivalent form by using the
Abel--Tauber theorem\cite{wiener,vanderpol}:
\begin{align}
\rho(t)  &  =\lim_{\epsilon\rightarrow0^{+}}\epsilon\int_{-\infty}%
^{0}ds\,e^{\epsilon s}\exp\left[  -i\mathbf{L}\left(  t-s\right)  \right]
\exp(-iL_{0}s)\Phi_{\lambda}\nonumber\\
&  =\exp\left(  -i\mathbf{L}t\right)  \lim_{\epsilon\rightarrow0^{+}}%
\epsilon\int_{-\infty}^{0}ds\,e^{\epsilon s}\exp\left(  i\mathbf{L}s\right)
\exp(-i\lambda s)\Phi_{\lambda}. \label{35}%
\end{align}
Physically, this form implies a time average of a wave train released from
remote past until $t=0$. We define a time-independent function $\psi_{\lambda
}^{(+)}$ by the formula
\begin{equation}
\psi_{\lambda}^{(+)}=\lim_{\epsilon\rightarrow0^{+}}\epsilon\int_{-\infty}%
^{0}ds\,e^{\epsilon s}\exp\left(  i\mathbf{L}s\right)  \exp(-i\lambda
s)\Phi_{\lambda} \label{36}%
\end{equation}
to formulate time-independent scattering theory, since it contains all the
necessary information on the scattering system. The nature of the scattered
function $\psi_{\lambda}^{(+)}$ can be better understood by recasting Eq.
(\ref{36}) on performing integration: with the definition of a complex number
\[
z=\lambda+i\epsilon
\]
we find
\begin{equation}
\psi_{\lambda}^{(+)}=-i\epsilon\left(  \mathbf{L}-z\right)  ^{-1}\Phi
_{\lambda}\equiv-i\epsilon\mathcal{R}\left(  z\right)  \Phi_{\lambda},
\label{37}%
\end{equation}
where $\ \mathcal{R}\left(  z\right)  $ is the resolvent operator and the
limit sign $\epsilon\rightarrow0^{+}$ is omitted for notational brevity. The
limit must be taken when the calculation of the right-hand side is completed.
This limit will be understood henceforth. Multiplying $\left(  \mathbf{L}%
-z\right)  $ to Eq. (\ref{37}) from left and taking the limit $\epsilon
\rightarrow0^{+}$, we obtain
\begin{equation}
\mathbf{L}\psi_{\lambda}^{(+)}=\lambda\psi_{\lambda}^{(+)}. \label{38}%
\end{equation}
This equation suggests that $\psi_{\lambda}^{(+)}$ is an eigenfunction of
$\mathbf{L}$ belonging to the same eigenvalue $\lambda$ as for $\mathbf{L}%
_{0}$. If it is recalled that the free state of the scattering system prepared
in the remote past also has the eigenvalue $\lambda$ with eigenfunction
$\Phi_{\lambda}$, we see that the full scattering eigenfunction $\psi
_{\lambda}^{(+)}$ and the free eigenfunction $\Phi_{\lambda}$ belong to the
same eigenvalue spectrum, that is, the scattering process occurs on the shell
of a given eigenvalue $\lambda$ of the Liouville operators and the total
eigenvalue of the system does not change even if there is scattering of
particles by each other. This is an important point worth keeping in mind when
we apply the classical scattering theory formalism to kinetic theory and, in
particular, to calculation of collision bracket integrals given in terms of
classical collision operators.

The classical resolvent operator in Eq. (\ref{37}) can be recast into another
alternative form
\begin{align}
\mathcal{R}\left(  z\right)   &  =\left(  \mathbf{L}-z\right)  ^{-1}%
\nonumber\\
&  =\mathcal{R}_{0}\left(  z\right)  -\mathcal{R}_{0}\left(  z\right)
\mathbf{L}_{1}\mathcal{R}\left(  z\right) \nonumber\\
&  =\mathcal{R}_{0}\left(  z\right)  -\mathcal{R}\left(  z\right)
\mathbf{L}_{1}\mathcal{R}_{0}\left(  z\right)  , \label{39}%
\end{align}
where the free resolvent operator $\mathcal{R}_{0}\left(  z\right)  $ is
defined by
\begin{equation}
\mathcal{R}_{0}\left(  z\right)  =\left(  \mathbf{L}_{0}-z\right)  ^{-1}.
\label{40}%
\end{equation}
Upon use of Eq. (\ref{39}), Eq. (\ref{37}) can be written in a more familiar
form
\begin{equation}
\psi_{\lambda}^{(+)}=\Phi_{\lambda}-\mathcal{R}_{0}\left(  z\right)
\mathbf{L}_{1}\psi_{\lambda}^{(+)}, \label{41}%
\end{equation}
which is called the classical Lippmann--Schwinger equation for the scattered
eigenfunction $\psi_{\lambda}^{(+)}$ in analogy to the quantum mechanical
Lippmann--Schwinger equation for scattering.\cite{newton,prugovecki} It is an
integral equation for $\psi_{\lambda}^{(+)}$. By defining the classical
collision operator $\mathbf{T}\left(  z\right)  $%
\begin{equation}
\mathbf{T}\left(  z\right)  =\mathbf{L}_{1}-\mathbf{L}_{1}\mathcal{R}%
_{0}\left(  z\right)  \mathbf{T}\left(  z\right)  , \label{42}%
\end{equation}
we may put the classical Lippmann-Schwinger equation in an alternative form
\begin{equation}
\psi_{\lambda}^{(+)}=\Phi_{\lambda}-\mathcal{R}_{0}\left(  z\right)
\mathbf{T}\left(  z\right)  \Phi_{\lambda}. \label{43}%
\end{equation}
The operator relations (\ref{39}) and (\ref{42}) can be rearranged to obtain
the relations
\begin{align}
\mathcal{R}\left(  z\right)  \mathbf{L}_{1}  &  =\mathcal{R}_{0}\left(
z\right)  \mathbf{T}\left(  z\right)  ,\label{44}\\
\mathbf{L}_{1}\mathcal{R}\left(  z\right)   &  =\mathbf{T}\left(  z\right)
\mathcal{R}_{0}\left(  z\right)  ,\label{44a}\\
\mathcal{R}\left(  z\right)   &  =\mathcal{R}_{0}\left(  z\right)
-\mathcal{R}_{0}\left(  z\right)  \mathbf{T}\left(  z\right)  \mathcal{R}%
_{0}\left(  z\right)  ,\label{45}\\
\mathbf{T}\left(  z\right)   &  =\mathbf{L}_{1}-\mathbf{L}_{1}\mathcal{R}%
\left(  z\right)  \mathbf{L}_{1}. \label{46}%
\end{align}
These relations will be found useful in various calculations involving
collision operators since the classical collision operators appear in the
theory of transport processes. They have been in fact used in some of formal
kinetic theories\cite{zwanzig,cohen,kawasaki} in the literature in 1960s and 1970s.

The equations presented remain structurally the same as for both two-particle
and many-particle collisions, and since they can be easily transcribed into
many-particle versions, we will simply consider the equations presented here
applicable to a many-particle situation. Since the many-particle collision
operators are often computed in terms of collision operators of a smaller
number of particles---usually two-particle operators---owing to the inherent
difficulty with many-particle dynamical problems by using some sort of
expansion, there are some precautions that must be taken when many-particle
collision and resolvent operators are calculated in terms of operators
involving smaller numbers of particles, for example, two-particle,
three-particle operators, and so on. For example, the often used binary
collision expansion\cite{zwanzig,cohen,kawasaki} for many-particle collision
operators is accompanied by a divergence difficulty, which can be avoided by
using a suitable cluster expansion\cite{eubook,cluster}. Since many-particle
operators are often decomposed into such lower-order operators of smaller
numbers of particles, the precautions mentioned become important when kinetic
theory results are computed for experimental comparison. Detailed discussions
on these points are given in Chapter 9 of Ref. 3 to which the interested
reader is referred.

\section{Alternative Forms for the Collision Bracket Integral\label{Sec4}}

In kinetic theory, either the kinetic equation is derived approximately from
the Liouville equation in a form breaking the time-reversal invariance or an
irreversible equation is postulated on the basis of the viewpoint that the
kinetic equation is a fundamental and irreversible equation for mesoscopic
description of a many-particle system, built on (reversible) classical or
quantum mechanics, not something that canbe derived from
time-reversal-invariant Hamiltonian equations of motion. The difference
between these approaches is just philosophical. Either way, we are treating
macroscopic systems from the molecular viewpoint by using an irreversible
equation, which is qualitatively different from the Newtonian equation of
motion or the Schr\"{o}dinger equation in the sense that the former is
time-reversal symmetry breaking whereas the latter is time-reversal invariant.
The time reversal symmetry is broken by the kinetic equation (e.g., Boltzmann
equation) because of the time-reversal symmetry-breaking collision term (e.g.,
the Boltzmann collision term) in which energy dissipation accompanying
irreversible processes is vested. The collision term is given in terms of the
collision cross section or collision operator which is relatively simple to
calculate in the case of two-particle elastic scattering, but becomes hard to
calculate as the complexity of the system increases if the number of particles
involved increases beyond two. All material functions such as transport
coefficients are given as collision-weighted averages of dynamical quantities,
and the theoretical formulas for material functions derived by any kinetic
theory cannot transcend the formalistic level to make connection with
experimental data unless a practicable method of computing collision-weighted
averages is developed. By method we do not mean some sort of analytical
theoretic method, but an algorithm by which the aforementioned averages can be
efficiently computed on a computer, for example, in a manner similar to the
molecular dynamic or Monte Carlo simulations for equilibrium fluids.
Therefore, the aim is to transform the formal expressions for material
functions in kinetic theory to relatively simple quadratures or forms readily
computable electronically. It is the main objective in this section and, in
fact, of this paper.

To the end stated, it is sufficient to consider the following collision
bracket integral appearing in the dense fluid kinetic theory formulated
elsewhere (Ref. 3):
\begin{equation}
I_{c}=i\left\langle h(x)\mathbf{T}\left(  z\right)  h(x)F_{eq}(x)\right\rangle
\equiv iJ_{c}, \label{47}%
\end{equation}
where $h(x)$ is a function of phase $x=\left(  \mathbf{p},\mathbf{r}\right)
\equiv(\mathbf{p}^{(N)},\mathbf{r}^{(N)})$ for an $N$ particle system; for
example, in the case of dilute gases%
\begin{equation}
h(x)=\sum_{j}m_{j}\left[  \mathbf{C}_{j}\mathbf{C}_{j}\right]  ^{(2)}%
,\;\frac{1}{2}\sum_{j}m_{j}C_{j}^{2}\mathbf{C}_{j}\mathbf{\quad}\left(
\mathbf{C}_{j}=\mathbf{v}_{j}-\mathbf{u}\right)  \label{47h}%
\end{equation}
with $\mathbf{v}_{j}$ and $\mathbf{u}$ denoting respectively the particle
velocity and the mean fluid velocity; $F_{eq}(x)$ is the equilibrium
distribution function; $T\left(  z\right)  $ is the $N$-particle collision
operator obeying the $N$-particle version of the classical Lippmann-Schwinger
equation (\ref{42}); and $i=\sqrt{-1}$. The first member on the right hand
side of Eq. (\ref{47h}) is for the shear stress and the second member is for
the heat flux in the case of a dilute monatomic gas. In the case of dense
fluids, $h(x)$ consists of virial tentors, energy flux tvectors, etc.; see
Ref. 3 for explicit forms for them and also Eq. (\ref{68}) for the virial
tensor below. The symbol $\left[  \mathbf{B}\right]  ^{(2)}$ stands for the
traceless symmetric part of second rank tensor $\mathbf{B}$. The angular
brackets stand for integration over the phase space. The limit $\epsilon
\rightarrow0^{+}$ must be taken on completion of calculation for the average
in Eq. (\ref{47}). We aim to cast the phase space average in Eq. (\ref{47}) in
an alternative, and more readily computable, form since it is not clear at
present how the classical collision operator $\mathbf{T}(z)$ may be
numerically evaluated on a computer.

It is convenient to use the bra and ket vector notation so that we can write
\begin{equation}
J_{c}=\left\langle g\mid\mathbf{T}\left(  z\right)  \mid g\right\rangle ,
\label{48}%
\end{equation}
where
\begin{equation}
\left\vert g\right\rangle =\left\vert h(x)F_{eq}^{1/2}\right\rangle .
\label{49}%
\end{equation}
Let $\Phi_{\lambda}$ denote the complete set of eigenfunctions for $L_{0}$,
the free Liouville operator for the $N$ particles. The complete set has the
closure
\begin{equation}
\sum_{\lambda}\left\vert \Phi_{\lambda}\left(  x\right)  \right\rangle
\left\langle \Phi_{\lambda}\left(  x\right)  \right\vert =\mathbf{1.}
\label{50}%
\end{equation}
The collision integral $J_{c}$ then can be written as
\begin{equation}
J_{c}=\sum_{\lambda}\sum_{\lambda^{\prime}}\left\langle g\left\vert
\Phi_{\lambda}\left(  x\right)  \right\rangle \left\langle \Phi_{\lambda
}\left(  x\right)  \right\vert \mathbf{T}\left(  z\right)  \left\vert
\Phi_{\lambda^{\prime}}\left(  x\right)  \right\rangle \left\langle
\Phi_{\lambda^{\prime}}\left(  x\right)  \right\vert g\right\rangle .
\label{51}%
\end{equation}
Upon rearranging Eq. (\ref{45}) into the following form
\begin{equation}
\mathbf{T}(z)=\mathcal{R}_{0}^{-1}\left(  z\right)  \left[  \mathcal{R}%
_{0}\left(  z\right)  -\mathcal{R}\left(  z\right)  \right]  \mathcal{R}%
_{0}^{-1}\left(  z\right)  \label{52}%
\end{equation}
and using it in Eq. (\ref{51}), the collision integral can be written as
\begin{align}
J_{c}  &  =\sum_{\lambda}\sum_{\lambda^{\prime}}\left(  \lambda-z\right)
^{\ast}\left(  \lambda^{\prime}-z\right)  \left\langle g\left\vert
\Phi_{\lambda}\left(  x\right)  \right\rangle \left\langle \Phi_{\lambda
}\left(  x\right)  \right\vert \mathcal{R}\left(  z\right)  \left\vert
\Phi_{\lambda^{\prime}}\left(  x\right)  \right\rangle \left\langle
\Phi_{\lambda^{\prime}}\left(  x\right)  \right\vert g\right\rangle
\nonumber\\
&  -\sum_{\lambda}\sum_{\lambda^{\prime}}\left(  \lambda^{\prime}-z\right)
\left\langle g\left\vert \Phi_{\lambda}\left(  x\right)  \right\rangle
\left\langle \Phi_{\lambda}\left(  x\right)  \mid\Phi_{\lambda^{\prime}%
}\left(  x\right)  \right\rangle \left\langle \Phi_{\lambda^{\prime}}\left(
x\right)  \right\vert g\right\rangle . \label{53}%
\end{align}
Since the eigenfunctions are orthogonal, the collision integral is further
simplified to the form
\begin{align}
J_{c}  &  =\sum_{\lambda}\sum_{\lambda^{\prime}}\left(  \lambda-z\right)
^{\ast}\left(  \lambda^{\prime}-z\right)  \left\langle g\left\vert
\Phi_{\lambda}\left(  x\right)  \right\rangle \left\langle \Phi_{\lambda
}\left(  x\right)  \right\vert \mathcal{R}\left(  z\right)  \left\vert
\Phi_{\lambda^{\prime}}\left(  x\right)  \right\rangle \left\langle
\Phi_{\lambda^{\prime}}\left(  x\right)  \right\vert g\right\rangle
\nonumber\\
&  -\sum_{\lambda}\left(  \lambda-z\right)  \left\langle g\left\vert
\Phi_{\lambda}\left(  x\right)  \right\rangle \left\langle \Phi_{\lambda
}\left(  x\right)  \right\vert g\right\rangle . \label{54}%
\end{align}
The scattering theory consideration made earlier in the previous section
indicates that the collision occurs on the shell of a given $\lambda$. This
means that there will be negligible contribution from the off-shell $\lambda$
and thus the important contribution is made from the term where
\[
\left(  \lambda-z\right)  ^{\ast}\left(  \lambda^{\prime}-z\right)  =-\left(
i\epsilon\right)  ^{2}=\epsilon^{2}.
\]
Therefore, we obtain
\begin{align}
J_{c}  &  =\epsilon^{2}\sum_{\lambda}\sum_{\lambda^{\prime}}\left\langle
g\left\vert \Phi_{\lambda}\left(  x\right)  \right\rangle \left\langle
\Phi_{\lambda}\left(  x\right)  \right\vert \mathcal{R}\left(  z\right)
\left\vert \Phi_{\lambda^{\prime}}\left(  x\right)  \right\rangle \left\langle
\Phi_{\lambda^{\prime}}\left(  x\right)  \right\vert g\right\rangle
\nonumber\\
&  \qquad-i\epsilon\sum_{\lambda}\left\langle g\left\vert \Phi_{\lambda
}\left(  x\right)  \right\rangle \left\langle \Phi_{\lambda}\left(  x\right)
\right\vert g\right\rangle \nonumber\\
&  =\epsilon^{2}\left\langle g\left\vert \mathcal{R}\left(  z\right)
\right\vert g\right\rangle -i\epsilon\left\langle g\mid g\right\rangle .
\label{54a}%
\end{align}
By using the definition of the resolvent operator the integral $J_{c}$ can be
written in the form of integral
\begin{equation}
J_{c}=i\epsilon\int_{0}^{\infty}dt\,e^{-\epsilon t}\left\langle e^{i\mathbf{L}%
t}g\mid g\right\rangle -i\epsilon\left\langle g\mid g\right\rangle .
\label{55}%
\end{equation}
Therefore, the collision integral $I_{c}$ may be written as
\begin{equation}
I_{c}=\epsilon\left[  \left\langle g\mid g\right\rangle -\epsilon\int
_{0}^{\infty}dt\,e^{-\epsilon t}\left\langle e^{iLt}g\mid g\right\rangle
\right]  . \label{56}%
\end{equation}
Since
\[
g(t)=e^{i\mathbf{L}t}g=F_{eq}^{1/2}e^{i\mathbf{L}t}h,
\]
owing to the fact that the equilibrium distribution function consists of
invariants of $\mathbf{L}$, we finally obtain
\begin{align}
I_{c}  &  =\epsilon\left[  \left\langle g\mid g\right\rangle -\epsilon\int
_{0}^{\infty}dt\,e^{-\epsilon t}\left\langle g(t)\mid g\right\rangle \right]
\nonumber\\
& \nonumber\\
&  =-\epsilon^{2}\int_{0}^{\infty}dt\,e^{-\epsilon t}\left\langle g(t)-g\mid
g\right\rangle . \label{57}%
\end{align}
This form and, particularly, the second term on the right in the first line,
is rather reminiscent of the time autocorrelation functions appearing in the
linear response theory\cite{green,kubo,mori}. Here $\epsilon$ may be taken
with the inverse of a sufficiently large value of time, $\tau_{c}$, so that
\begin{align}
I_{c}  &  =-\tau_{c}^{-2}\int_{0}^{\infty}dt\,e^{-t/\tau_{c}}\left\langle
g(t)-g\mid g\right\rangle \nonumber\\
&  =\tau_{c}^{-1}\int_{0}^{\infty}ds\,e^{-s}\left\langle \Delta g(s\tau
_{c})\mid g\right\rangle \quad\left(  s=t/\tau_{c}\right)  , \label{57a}%
\end{align}
where%
\begin{equation}
\Delta g\left(  s\tau_{c}\right)  =g-g(s\tau_{c}). \label{57b}%
\end{equation}
This is an alternative form for the collision bracket integral in Eq.
(\ref{47}). It is given in terms of a time correlation function which is
certainly more readily amenable to molecular dynamic simulation methods than
the one involving the classical $N$-particle collision operator $\mathbf{T}%
(z).$

If there exists a plateau value region in the integrand, $\Delta g\left(
s\tau_{c}\right)  =g-g(s\tau_{c})$ may be approximated by $\Delta g\left(
\tau_{c}\right)  $, the integral may be approximated by
\begin{equation}
I_{c}=\frac{1}{\tau_{c}}\left\langle \Delta g\left(  \tau_{c}\right)  \mid
g\right\rangle , \label{59}%
\end{equation}
By reversing the collision process in the phase space and using the time
reversal invariance of the phase volume, it is possible to put this integral
in a symmetric form
\begin{align}
I_{c}  &  =\frac{1}{2\tau_{c}}\left\langle \Delta g\left(  \tau_{c}\right)
\mid\Delta g\left(  \tau_{c}\right)  \right\rangle \nonumber\\
&  =\frac{1}{2\tau_{c}}\int dx^{(N)}\Delta g\left(  \tau_{c}\right)  \Delta
g\left(  \tau_{c}\right)  F_{eq}^{(N)}\left(  x^{(N)}\right)  , \label{58}%
\end{align}
which is reminiscent of the collision bracket integrals in the Chapman-Enskog
theory\cite{chapman,ferziger} of dilute gases, but this form is not limited to
dilute gases. This is another alternative form for the collision bracket
integral we set out to show in this paper.

This integral indeed can be shown to be the Chapman-Enskog collision bracket
integral for $\Delta g\left(  \tau_{c}\right)  $ if we consider the case of
two particles for $I_{c}.$ For this purpose, let us write $I_{c}$ explicitly
for two particles:
\begin{equation}
I_{c}=\frac{1}{2\tau_{c}}\int dx^{(2)}\Delta g\left(  \tau_{c}\right)  \Delta
g\left(  \tau_{c}\right)  F_{eq}^{(2)}\left(  x^{(2)}\right)  . \label{60}%
\end{equation}
If it is assumed that $\Delta g\left(  \tau_{c}\right)  $ depends only on the
momenta and $F_{eq}^{(2)}\left(  x^{(2)}\right)  =f_{eq}(\mathbf{p}_{1}%
)f_{eq}(\mathbf{p}_{2})$ as is the case of the Chapman-Enskog collision
bracket integrals, then it is possible to show$^{2,3,14}$ that
\begin{equation}
\int dx^{(2)}\cdots=\tau_{c}V\int d\mathbf{v}_{1}\int d\mathbf{v}_{2}\int
_{0}^{2\pi}d\varphi\int_{0}^{\infty}db\,bg_{12}\cdots\label{61}%
\end{equation}
in the standard notation. Here $V$ is the volume of the container, $b$ is the
impact parameter, $\varphi$ is the azimuthal scattering angle, $g_{12}$ is the
relative speed, and $f_{eq}(\mathbf{p}_{i})$ is a singlet equilibrium momentum
distribution function. By using this result, we finally get
\begin{equation}
I_{c}^{(2)}=\frac{1}{2}V\int d\mathbf{v}_{1}\int d\mathbf{v}_{2}\int_{0}%
^{2\pi}d\varphi\int_{0}^{\infty}db\,bg_{12}\Delta g\left(  \tau_{c}\right)
\Delta g\left(  \tau_{c}\right)  f_{eq}(\mathbf{p}_{1})f_{eq}(\mathbf{p}_{2}),
\label{62}%
\end{equation}
which, apart from the constant factor, is clearly the Chapman-Enskog collision
bracket integral\cite{chapman,ferziger} for $\Delta g\left(  \tau_{c}\right)
=g-g\left(  \tau_{c}\right)  $ where $g\left(  \tau_{c}\right)  $ is the
post-collision value of $g$ if $\tau_{c}$ is taken for the collision time for
the particles.

In the case of $I_{c}$ involving three particles, we employ, for example,
mass-normalized coordinates\cite{note} which are subsequently expressed in
hyperspherical coordinates\cite{erdelyi}. In the aforementioned
mass-normalized hyperspherical coordinates the three-particle phase integral
can be written as\cite{eubook,euthree}
\begin{equation}
\int dx^{(3)}=\tau_{c}V\int d\mathbf{P}_{c}\int d%
\mbox{\boldmath$\pi$}%
_{1}\int d%
\mbox{\boldmath$\pi$}%
_{2}\int d\Omega_{4}\int_{0}^{\infty}db\,b^{4}\left(  P/\mu\right)  ,
\label{63}%
\end{equation}
where $\mathbf{P}_{c}$ is the center of mass momentum, $%
\mbox{\boldmath$\pi$}%
_{1}$ and $%
\mbox{\boldmath$\pi$}%
_{2}$ are two components of mass-normalized relative momenta of the three
particles, $P=\left\vert
\mbox{\boldmath$\pi$}%
_{1}+%
\mbox{\boldmath$\pi$}%
_{2}\right\vert $, $\mu^{2}=m_{1}m_{2}m_{3}/\left(  m_{1}+m_{2}+m_{3}\right)
=m^{2}/3$,
\begin{equation}
d\Omega_{4}=\sin^{3}\theta_{4}\sin^{2}\theta_{3}\sin\theta_{2}d\theta
_{1}d\theta_{2}d\theta_{3}d\theta_{4} \label{64}%
\end{equation}
with $\theta_{2},$ $\theta_{3},$ $\theta_{4}$ denoting the hyperpolar angles
$\left(  0\leq\theta_{i}\leq\pi:i=2,3,4\right)  $, $\theta_{1}$ is the
hyperazimuthal angle $\left(  0\leq\theta_{1}\leq2\pi\right)  $, and $b$ is
the generalized impact parameter. For details of the coordinate transformation
and the collision dynamics leading to Eqs. (\ref{63}) and (\ref{64}), see Sec.
13.6 of Ref. \cite{eubook}. By using this result, the three-particle collision
bracket integral can be written as
\begin{align}
I_{c}^{(3)}  &  =\frac{1}{2}V\int d\mathbf{P}_{c}\int d%
\mbox{\boldmath$\pi$}%
_{1}\int d%
\mbox{\boldmath$\pi$}%
_{2}\int d\Omega_{4}\int_{0}^{\infty}db\,b^{4}\left(  P/\mu\right)  \Delta
g\left(  \tau_{c}\right)  \Delta g\left(  \tau_{c}\right) \nonumber\\
&  \qquad\qquad\qquad\times f_{eq}(%
\mbox{\boldmath$\pi$}%
_{1})f_{eq}(%
\mbox{\boldmath$\pi$}%
_{2})f_{eq}(\mathbf{P}_{c}), \label{65}%
\end{align}
in the case where $\Delta g\left(  \tau_{c}\right)  $ depends on the momenta
of three particles only and
\[
F_{eq}^{(3)}=f_{eq}(%
\mbox{\boldmath$\pi$}%
_{1})f_{eq}(%
\mbox{\boldmath$\pi$}%
_{2})f_{eq}(\mathbf{P}_{c}).
\]
It must be noted that parameter $\tau_{c}$ is canceled out in $I_{c}^{(3)}$ if
$\tau_{c}$ is taken as the collision time of the three particles. Here we
remark that in the case of $N$-particle collisions $I_{c}$ in Eq. (\ref{60}),
that is, $\mathbf{T}\left(  z\right)  $, may be expanded into a cluster
expansion\cite{eubook} where $I_{c}^{(2)}$ and $I_{c}^{(3)}$ appear as the
leading order contributions in a density series. Since such an expansion is
not the aim of this work, we will not discuss it in this paper.

\section{Concluding Remarks\label{Sec5}}

In this paper, we have re-examined the eigenvalue problem of the Liouville
operator in the context of kinetic theory. The eigenfunctions of the Liouville
operator are shown possible to construct by making use of canonical
transformation. The eigenfunctions obtained enable us to consider classical
scattering theory in a mathematically precise manner. An important point
realized in this investigation is that the classical scattering involves waves
moving perpendicularly to the characteristic function $W$ with phases given by
Eq. (\ref{24}) and occurs on the shell of a fixed eigenvalue which is shared
by both the incident and scattered waves in the phase space. By using this
latter notion in the calculation of the collision bracket integrals appearing
in kinetic theory of dense fluids, it is possible to recast them in terms of a
time correlation function. This latter form can be put in an approximate form
reminiscent of the collision bracket integral in the Chapman-Enskog theory of
dilute gases, if the there exists a plateau value region in the dynamic
quanties involved in the time correlation functions. Otherwise, the collision
bracket integrals are expressed in the form of a time integral of time
correlation functions of $\Delta g\left(  s\tau_{c}\right)  $ averaged over
the equilibrium ensemble distribution functions.

The recast forms for the collision bracket integral are certainly more
suitable for numerical computation on a computer than the form given in terms
of the classical collision operator $\mathbf{T}\left(  z\right)  $. The
two-particle example is shown to give rise to the traditional Chapman-Enskog
theory result for dilute monatomic gases. We have also presented a
three-particle collision bracket integral in the case of dilute gases where
there is no statistical correlation. The alternative forms of the collision
bracket integral present a possibility of developing a new way of computing
many-particle collision bracket integrals on a computer. With this
investigation we now have achieved a numerical algorithm to compute the
transport coefficients appearing the kinetic theory of dense fluids that is
comparable to that of linear response theory.

Finally, we note here the relation of the collision bracket integral to the
transport coefficient. For example, the viscosity is related to the collision
bracket integrals in the following manner\cite{eubk4}:%
\begin{equation}
\eta=\frac{5m}{16p^{2}}\frac{1}{\mathbb{R}^{(1,1)}}, \label{66}%
\end{equation}
where $\sigma$ is the diameter of the molecule, $m$ is the reduced mass, $n$
is the number density, and%
\begin{align}
\mathbb{R}^{(1,1)}  &  =\int dx^{(N)}F_{eq}^{(N)}\left(  x^{(N)}\right)
\sum_{j,k}^{N}\delta\left(  \mathbf{r}_{j}-\mathbf{r}\right)  h_{j}%
^{(1)}:\mathbf{T}^{(N)}\left(  z\right)  h_{k}^{(1)}\nonumber\\
&  =\tau_{c}^{-1}\int_{0}^{\infty}ds\left\langle \Delta h\left(  s\tau
_{c}\right)  |\Delta h\left(  s\tau_{c}\right)  \right\rangle ,\label{67}\\
\Delta h\left(  s\tau_{c}\right)   &  =\sum_{j}\left[  h_{j}^{(1)}\left(
s\tau_{c}\right)  -h_{j}^{(1)}\left(  0\right)  \right]  \label{67b}%
\end{align}
with $h_{j}^{(1)}$ standing for the virial tensor%
\begin{align}
h_{j}^{(1)}  &  =\left[  mC_{j}C_{j}\right]  ^{(2)}-\frac{1}{2}\sum_{j\neq
l}\frac{u_{jl}^{\prime}}{r_{jl}}\left[  \mathbf{r}_{jl}\mathbf{r}_{jl}\right]
^{(2)}\label{68}\\
u_{jl}^{\prime}  &  =\frac{\partial u_{jl}\left(  r_{jl}\right)  }{\partial
r_{jl}}\qquad r_{jl}=\left\vert \mathbf{r}_{j}-\mathbf{r}_{l}\right\vert
,\quad u_{jl}=\text{ potential energy}\label{69}\\
\left[  \mathbf{A}\right]  ^{(2)}  &  =\frac{1}{2}\left(  \mathbf{A}%
+\mathbf{A}^{t}\right)  -\frac{1}{3}\mathbf{\delta}\text{Tr}\mathbf{A}\text{,
traceless symmetric part of tensor }\mathbf{A.} \label{70}%
\end{align}
which is in the form, namely, the collision bracket integral considered in the
previous section. The collision bracket integral on the right can be expressed
in the alternative form, Eq. (\ref{58}), presented in the previous section.
Thus the viscosity is inversely proportional to the collision bracket
integral. Separating the center of mass part of $dx^{(N)}$ so that
$dx^{(N)}=d\mathbf{r}d\mathbf{P}dx^{(N/c)}$, $\left\langle \Delta h\left(
s\tau_{c}\right)  |\Delta h\left(  s\tau_{c}\right)  \right\rangle $ may be
cast into the form%
\begin{equation}
\left\langle \Delta h\left(  s\tau_{c}\right)  |\Delta h\left(  s\tau
_{c}\right)  \right\rangle =\int d\mathbf{P}\int dx^{(N/c)}F_{eq}%
^{(N/c)}\left(  x^{(N/c)}\right)  \Delta h\left(  s\tau_{c}\right)  :\Delta
h\left(  s\tau_{c}\right)  , \label{71}%
\end{equation}
where $F_{eq}^{(N/c)}$ is the equilibrium distribution function excluding the
center of mass motion part of the $N$ particles.

Other transport coefficients may be expressed similarly in terms of the
collision bracket integrals considered earlier in this work. It should be
noted that the manner in which the time correlation function Eq. (\ref{58})
appears in the transport coefficient and also the meaning of the time
correlation function are different from those in the linear response theory.
This difference stems from the fact that the present kinetic theory and the
linear response theory differ in their basic approach to transport processes.

\textbf{Acknowledgment}

This work has been supported in part by the Discovery grants from the Natural
Sciences and Engineering Research Council of Canada.

\appendix{}

\section{Some Mathematical Properties of Eigenfunctions and Eigenvalues}

The Liouville operator $\mathbf{L}$ may have eigenfunctions and eigenvalues,
but also has some important mathematical properties, which are useful and
necessary for developing physical theory with it. We list them in the following.

\begin{description}
\item[Property 1.] The Liouville operator is hermitean. Consider%
\[
l_{\lambda\lambda^{\prime}}=\int dx\psi_{\lambda^{\prime}}^{\ast}%
\mathbf{L}\psi_{\lambda}.
\]
Then%
\begin{align*}
l_{\lambda\lambda^{\prime}}^{\ast}  &  =\int dx\left(  \psi_{\lambda^{\prime}%
}^{\ast}\mathbf{L}\psi_{\lambda}\right)  ^{\ast}\\
&  =\int dx\psi_{\lambda^{\prime}}\mathbf{L}^{\ast}\\
&  =\int dx\psi_{\lambda^{\prime}}i\left[  H,\psi_{\lambda}^{\ast}\right]
_{pq}\\
&  =\int dx\psi_{\lambda}^{\ast}\left(  -i\right)  \left[  H,\psi
_{\lambda^{\prime}}\right]  _{pq}\\
&  =\int dx\psi_{\lambda}^{\ast}\mathbf{L}\psi_{\lambda^{\prime}}\\
&  =l_{\lambda^{\prime}\lambda}%
\end{align*}
Therefore $\mathbf{L}$ is hermitean and hence
\begin{equation}
\mathbf{L}^{\dag}=\mathbf{L.} \label{A1}%
\end{equation}

\item[Propertiy 2.] Since $\mathbf{L}$ is hermitean, its eigenvalues are real.
Let%
\[
\mathbf{L}\psi_{\lambda}=\lambda\psi_{\lambda}.
\]
Then%
\[
\mathbf{L}^{\ast}\psi_{\lambda}^{\ast}=\lambda^{\ast}\psi_{\lambda}^{\ast}.
\]
Hence it follows that%
\begin{align*}
\int dx\left(  \psi_{\lambda}^{\ast}\mathbf{L}\psi_{\lambda}-\psi_{\lambda
}\mathbf{L}^{\ast}\psi_{\lambda}^{\ast}\right)   &  =\left(  \lambda
-\lambda^{\ast}\right)  \int dx\psi_{\lambda}^{\ast}\psi_{\lambda}\\
&  =\lambda-\lambda^{\ast}%
\end{align*}
But by virtue of hermiticity%
\[
\int dx\left(  \psi_{\lambda}^{\ast}\mathbf{L}\psi_{\lambda}-\psi_{\lambda
}\mathbf{L}^{\ast}\psi_{\lambda}^{\ast}\right)  =0.
\]
Therefore%
\begin{equation}
\lambda=\lambda^{\ast}, \label{A2}%
\end{equation}
which implies $\lambda$ is real.

\item[Property 3.] The eigenfunctions belonging to different eigenvalues are
orthogonal. Proof for this follows from Property 2.

\item[Property 4.] Any function in the phase space can be expanded into the
eigenfunctions of $\mathbf{L}$.%
\begin{equation}
f\left(  \mathbf{p},\mathbf{q}\right)  =\sum_{k}a_{k}\psi_{k}\left(
\mathbf{p},\mathbf{q}\right)  , \label{A3}%
\end{equation}
where%
\[
a_{k}=\int d\Gamma f\left(  \mathbf{p},\mathbf{q}\right)  \psi_{k}^{\ast
}\left(  \mathbf{p},\mathbf{q}\right)  .
\]
This property follows from the fact that $\mathbf{L}$ is a linear operator. In
the case of time-dependent functions%
\begin{equation}
\rho\left(  \mathbf{p},\mathbf{q,t}\right)  =\sum_{k}a_{k}\left(  t\right)
\psi_{k}\left(  \mathbf{p},\mathbf{q}\right)  . \label{A4}%
\end{equation}
Inserting it into the Liouville equation yields%
\begin{align*}
i\frac{\partial\rho}{\partial t}  &  =\sum_{k}a_{k}\left(  t\right)
\mathbf{L}\psi_{k}\left(  \mathbf{p},\mathbf{q}\right) \\
&  =\sum_{k}a_{k}\left(  t\right)  \lambda_{k}\psi_{k}\left(  \mathbf{p}%
,\mathbf{q}\right) \\
&  =i\sum_{k}\frac{da_{k}}{dt}\psi_{k}\left(  \mathbf{p},\mathbf{q}\right)  ,
\end{align*}
which gives rise to the ordinary differential equation%
\begin{equation}
i\frac{da_{k}}{dt}=\lambda_{k}a_{k}\left(  t\right)  . \label{A5}%
\end{equation}
Integrating this we obtain%
\[
a_{k}\left(  t\right)  =c_{k}\exp\left(  -i\lambda_{k}t\right)
\]
Finally,%
\begin{equation}
\rho\left(  \mathbf{p},\mathbf{q,}t\right)  =\sum_{k}c_{k}e^{-i\lambda_{k}%
t}\psi_{k}\left(  \mathbf{p},\mathbf{q}\right)  . \label{A6}%
\end{equation}
Of course, the eigenvalues and eigenfunctions appearing here are not known in
terms of the system properties. Nevertheless, it is possible to deduce some
general properties of the expansion, eigenvalues, and eigenfunctions, and they
may help assess eigenvalues and eigenfunctions we will obtain later.

\item[Property 5.] Since the distribution function is normalized,
\begin{equation}
\int dx\rho\left(  \mathbf{p},\mathbf{q,}t\right)  =1, \label{A7}%
\end{equation}
it follows%
\[
\sum_{k}c_{k}e^{-i\lambda_{k}t}\int dx\psi_{k}\left(  \mathbf{p}%
,\mathbf{q}\right)  =1.
\]
Since%
\[
\int dx\mathbf{L}\psi_{k}\left(  \mathbf{p},\mathbf{q}\right)  =\lambda
_{k}\int dx\psi_{k}\left(  \mathbf{p},\mathbf{q}\right)
\]
but%
\begin{align*}
\int dx\mathbf{L}\psi_{k}\left(  \mathbf{p},\mathbf{q}\right)   &  =-i\sum
_{j}\int dx\left(  \frac{\partial H}{\partial p_{j}}\frac{\partial\psi_{k}%
}{\partial q_{j}}-\frac{\partial H}{\partial q_{j}}\frac{\partial\psi_{k}%
}{\partial p_{j}}\right) \\
&  =i\sum_{j}\int dx\left(  \frac{\partial H}{\partial q_{j}\partial p_{j}%
}-\frac{\partial H}{\partial p_{j}\partial q_{j}}\right)  \psi_{k}\\
&  -i\sum_{j}\int dx\left(  \frac{\partial H}{\partial p_{j}}\psi_{k}%
|_{q_{j}=boundary}-\frac{\partial H}{\partial q_{j}}\psi_{k}|_{p_{j}%
=boundary}\right) \\
&  =0
\end{align*}
if%
\[
\frac{\partial H}{\partial p_{j}}\psi_{k}|_{q_{j}=boundary}=\frac{\partial
H}{\partial q_{j}}\psi_{k}|_{p_{j}=boundary}=0,
\]
it follows%
\begin{equation}
\lambda_{k}\int dx\psi_{k}\left(  \mathbf{p},\mathbf{q}\right)  =0. \label{A8}%
\end{equation}
This means%
\begin{equation}
\int dx\psi_{k}\left(  \mathbf{p},\mathbf{q}\right)  =0 \label{A9}%
\end{equation}
if $\lambda_{k}\neq0$. However, $\lambda_{k}\neq0$ is not equal to zero for
$k=0$. Therefore the integral of eigenfunctions for $k\neq0$ vanish. It thus
follows%
\[
c_{0}\int dx\psi_{0}\left(  \mathbf{p},\mathbf{q}\right)  =1.
\]
That is,
\begin{equation}
c_{0}=1/\int dx\psi_{0}\left(  \mathbf{p},\mathbf{q}\right)  \label{A10}%
\end{equation}
and the normalization of $\rho\left(  \mathbf{p},\mathbf{q,}t\right)  $ is
fulfilled.\newline In this manner, the distribution function in the nuclear
space---the phase space---is expandable in the eigenfunctions and its
expansion is reconciled with the notion of eigenfunctions being $\mathcal{L}%
^{2}$ functions in the Hilbert space.

\item[Property 6] It is well known that the Liouville operator is invariant to
time reversal times complex complex conjugation. That is, if we denote by
$\theta$ the time reversal operator then%
\[
\theta\mathbf{L=}-\mathbf{L}.
\]
Therefore if we denote the compound operator of time reversal and complex
conjugation by $\vartheta$
\[
\vartheta=\theta\left(  ^{\ast}\right)  =\left(  ^{\ast}\right)  \theta
\]
then
\[
\vartheta\mathbf{L}=\mathbf{L}%
\]
Therefore the time reversed distribution function $\rho\left(  -\mathbf{p}%
,\mathbf{q,}-t\right)  $ obeys the same Liouville equation as $\rho\left(
\mathbf{p},\mathbf{q,}t\right)  $. We note that this is in the root cause of
the difficulty in kinetic theory of matter when we develop it by means of the
Liouville equation, which is time reversal invariant. In connection with this
operator $\vartheta$, we note the following: Since
\[
\mathbf{L}^{\ast}=-\mathbf{L}%
\]
and the eigenvalues are real we have%
\begin{equation}
\mathbf{L}^{\ast}\psi_{\lambda_{k}}^{\ast}=\lambda_{k}\psi_{\lambda_{k}}%
^{\ast} \label{ceigeq}%
\end{equation}
and hence%
\[
\mathbf{L}\psi_{\lambda_{k}}^{\ast}=-\lambda_{k}\psi_{\lambda_{k}}^{\ast}.
\]
This means that $\psi_{\lambda_{k}}^{\ast}$ is the eigenfunction belonging to
$-\lambda_{k}$. Furthermore, since%
\[
\theta^{2}=1,
\]
it also follows%
\[
\theta\mathbf{L}^{\ast}=-\theta\mathbf{L}=\mathbf{\vartheta\mathbf{L}%
}=\mathbf{L.}%
\]
Now operating $\theta$ on Eq. (\ref{ceigeq}) we obtain%
\[
\theta\mathbf{L}^{\ast}\psi_{\lambda_{k}}^{\ast}=\lambda_{k}\theta
\psi_{\lambda_{k}}^{\ast}%
\]
but%
\[
\theta\mathbf{L}^{\ast}\psi_{\lambda_{k}}^{\ast}=\mathbf{L}\theta\psi
_{\lambda_{k}}^{\ast}%
\]
and hence%
\[
\mathbf{L}\theta\psi_{\lambda_{k}}^{\ast}=\lambda_{k}\theta\psi_{\lambda_{k}%
}^{\ast},
\]
which implies that%
\[
\theta\psi_{\lambda_{k}}^{\ast}=\text{ an eigenfunction of }\mathbf{L}\text{
belonging to }\lambda_{k}\equiv\psi_{\lambda_{k}}\text{.}%
\]
That is,
\begin{equation}
\psi_{\lambda_{k}}^{\ast}=\theta\psi_{\lambda_{k}}. \label{eigfrev}%
\end{equation}
This property is important for proving the orthogonality of eigenfunctions.

\item[Property 7.] The eigenfunctions $\psi_{\lambda}$ are generally complex
and hence they can be written in terms of amplitude and phase:%
\begin{equation}
\psi_{\lambda}\left(  \mathbf{p,q}\right)  =A_{\lambda}\left(  \mathbf{p,q}%
\right)  \exp\left[  i\Gamma_{\lambda}\left(  \mathbf{p,q}\right)  \right]
\label{A&P}%
\end{equation}
where%
\begin{align}
\mathbf{L}A_{\lambda}\left(  \mathbf{p,q}\right)   &  =0,\label{A}\\
\mathbf{L}\Gamma_{\lambda}\left(  \mathbf{p,q}\right)   &  =-i\lambda.
\label{P}%
\end{align}
This easily follows from the eigenvalue problem of $\mathbf{L}$ on
substituting Eq. (\ref{A&P}). Eq. (\ref{A}) implies $A_{\lambda}\left(
\mathbf{p,q}\right)  $ is made up of invariants of motion and $\Gamma
_{\lambda}\left(  \mathbf{p,q}\right)  $\ is the phase of a wave in the
$\Gamma$ space that corresponds to the eigenvalue $\lambda$. The eigenvalue
problem of $\mathbf{L}$ is now reduced to finding the amplitude and the phase
of waves. It is done in the main text.
\end{description}

\end{document}